\documentclass[a4paper,11pt]{spieman} 
\usepackage{amsmath,amsfonts,amssymb}
\usepackage{graphicx}
\usepackage{setspace}
\usepackage{textcomp}
\usepackage[version=4]{mhchem}
\usepackage[dvipsnames]{xcolor}

\title{Nanophotonic Light Management for \\Perovskite-Silicon Tandem Solar Cells}

\author[a,b]{Duote~Chen}
\author[a,b]{Phillip~Manley}
\author[a]{Philipp~Tockhorn}
\author[a]{David~Eisenhauer}
\author[a]{Grit~K\"{o}ppel}
\author[b,c]{Martin~Hammerschmidt}
\author[b,c]{Sven~Burger}
\author[a]{Steve~Albrecht}
\author[a]{Christiane~Becker}
\author[a,b,*]{Klaus~J\"{a}ger}
\affil[a]{Helmholtz-Zentrum Berlin f\"{u}r Materialien und Energie GmbH, Kekul\'{e}stra\ss e 5, 12489 Berlin, Germany}
\affil[b]{Zuse Institute Berlin, Takustra\ss e 7, 14195 Berlin, Germany}
\affil[c]{JCMwave GmbH, Bolivarallee 22, 14050 Berlin, Germany}

\begin{document} 
\maketitle

\begin{abstract}
Perovskite-silicon tandem solar cells are currently one of the most investigated concepts to overcome the theoretical limit for the power conversion efficiency of silicon solar cells. For monolithic tandem solar cells the available light must be distributed equally between the two subcells, which is known as current matching. For a planar device design, a global optimization of the layer thicknesses in the perovskite top cell allows current matching to be reached and reflective losses of the solar cell to be minimized at the same time. However, even after this optimization reflection and parasitic absorption losses occur, which add up to 7~mA/cm$^{2}$.

In this contribution we use numerical simulations to study, how well hexagonal sinusoidal nanotextures in the perovskite top-cell can reduce the reflective losses of the combined tandem device. We investigate three configurations. The current density utilization can be increased from 91\% for the optimized planar reference to 98\% for the best nanotextured device (period 500~nm and peak-to-valley height 500~nm), where 100\% refers to the Tiedje-Yablonovitch limit. In a first attempt to experimentally realize such nanophotonically structured perovskite solar cells for monolithic tandems, we investigate the morphology of perovskite layers, which are deposited onto sinusoidally structured substrates.
\end{abstract}


{\noindent \footnotesize\textbf{*}Klaus J\"{a}ger,  \linkable{klaus.jaeger@helmholtz-berlin.de} }

\section{Introduction}

Crystalline silicon (c-Si) solar cells have achieved enormous dominance in the photovoltaic market due to their high efficiencies, excellent material quality and continual reduction in manufacturing costs. The current record efficiency of 26.7\% \cite{green:2018eff51} is already 89\% of the theoretical limit value for the power conversion efficiency for single-junction solar cells \cite{richter:2013}. Further improvement of this technology will become more and more difficult. However, to further boost the distribution of photovoltaics in the future, higher efficiencies at low costs are required \cite{green:2016}.

We aim to surpass the Shockley-Queisser limit \cite{shockley:1961}---the most important technological limit for single-junction solar cells---with tandem solar cells, which combine two solar cells with different bandgaps. The incident sunlight first hits the top cell which has a higher bandgap and harvests the high-energy photons at a higher voltage, while the low-energy photons are transferred to the bottom cell which has a lower bandgap and corresponding lower voltage. In this way high-energy photons are able to contribute more voltage to the device instead of losing their excess energy by thermalization.

Lead halide perovskite materials have excellent optical properties for tandem applications due to their steep absorption edge \cite{dewolf:2014} and a tunable bandgap \cite{eperon:2014, beal:2016}. As a result, tandem cells with a perovskite top cell and a silicon bottom cell have the potential to reach efficiencies beyond 30\% \cite{filipic:2015, lal:2014,grant:2016}, if optical losses are adequately adressed \cite{santbergen:2016}.

In monolithic (two-terminal) tandem cells, the perovskite top cell and the c-Si bottom cell are electrically connected in series. Hence, for high efficiencies the photocurrent density of the top and bottom cell must be matched. We recently numerically maximized the achievable photocurrent density for different planar monolithic perovskite-silicon tandem device architectures with anti-reflective coatings by optimizing the layer thicknesses. The architecture with the electron-selective contact and $n$-doped layers on the front of the perovskite and silicon cells, respectively \cite{bush:2017}, enables efficiencies exceeding 30\%, when the perovskite bandgap is optimized as well\cite{jaeger:2017pero}. However, for fully planar devices, reflection losses cause a significant limitation for the matched photocurrent density $J_\text{ph}$.

Textured interfaces can reduce reflection losses via (1) enhancing coupling of light into the structure and (2) scattering light such that the average path length in the absorber is increased, leading to increased absorption especially in weakly absorbing regimes. In recent years, much effort has been put into developing 2-dimensional (2D) structures for light trapping \cite{gjessing:2011,grandidier:2011,battaglia:2012, mokkapati:2012, spinelli:2013, sprafke:2013, eisenlohr:2014,ha:2016}. For nanostructures, which texture the electrically active layers of the solar cell, it is mandatory that they have no detrimental effect on the electric solar cell performance. Therefore, we decided to focus on \emph{hexagonal sinusoidal nanotextures},\cite{jaeger:2016opex, jaeger:2018opex} which allow to combine a strong anti-reflective effect with good electrical performance, as we demonstrated for liquid-phase crystallized silicon thin-film solar cells.\cite{koeppel:2016, koeppel:2017}

In this work, we investigate, how sinusoidal nanotextures affect the optical performance of perovskite-silicon tandem solar cells. In order to efficiently optimize the perovskite layer thickness for current matching, we use Newton's method, which is based on linear approximation.  As a motivation, we demonstrate  the experimental feasibility of spin-coating perovskite layers onto sinusoidally nanotextured substrates.

\section{Experimental motivation}
\label{sec:exp}
 
So far, little experimental evidence of nanostructured perovskite layers was published: in some studies, different nanostructures were imprinted into the perovskite top surface.\cite{paetzold:2015apl,brittman:2017,pourdavoud:2017} While these top surface patterning techniques demand post-processing of the perovskite film after fabrication, other studies report on perovskite grown onto different kinds of nanophotonically patterned charge selective layers\cite{lin:2016,lee:2015}.

Here, we demonstrate, how sinusoidally nanotextured films can be fabricated with \emph{spin coating}. This deposition method is chosen, since it yields the best performing perovskite absorber layers for photovoltaic application so far. As model systems we use two kinds of glass substrates, which are covered with sinusoidal nanotextures.  The periods are $P = 500$~nm and $P = 750$~nm; the respective valley-to-peak heights are $h = 80$~nm and $h = 200$~nm, which correspond to aspect ratios of $a=0.16$ and $a=0.27$, respectively. The nanotextures were manufactured with nano-imprint lithography using a hybrid polymer with glass-like properties (OrmoComp by micro resist technology GmbH). Details on our nano-imprint lithography process can be found in Ref.\ \citenum{becker:2014}.

For better imaging capabilities, we thermally evaporate 30~nm of silver onto the nanotextured substrates. In addition, we prepare a flat reference substrate by spin-coating a thin layer of polytriarylamine (PTAA) onto a indium tin oxide (ITO) substrate. PTAA is a common hole transporting material (HTM) in p-i-n perovskite solar cells \cite{jost:2017,wolff:2017,stolterfoht:2017} which supports the uniform growth of perovskite.

For the perovskite layer fabrication in our experiment, we use the mixed halide perovskite composition \ce{Cs_{0.05}(FA_{0.83}MA_{0.17})_{0.95}Pb(I_{0.83}Br_{0.17})_3} using a recipe  by Saliba et al.\cite{Saliba:2016}.

We spin-coated the as-prepared 0.8~M solution according to the antisolvent method to promote the crystallization into perovskite. This method enables crystallization starting from the top interface of the perovskite\cite{minemawari:2011} film, which allows to grow films with a high optoelectronic quality, which is rather independent of the (nanotextured) substrate\cite{kegelmann:2017}.

\begin{figure}
\centering
\includegraphics{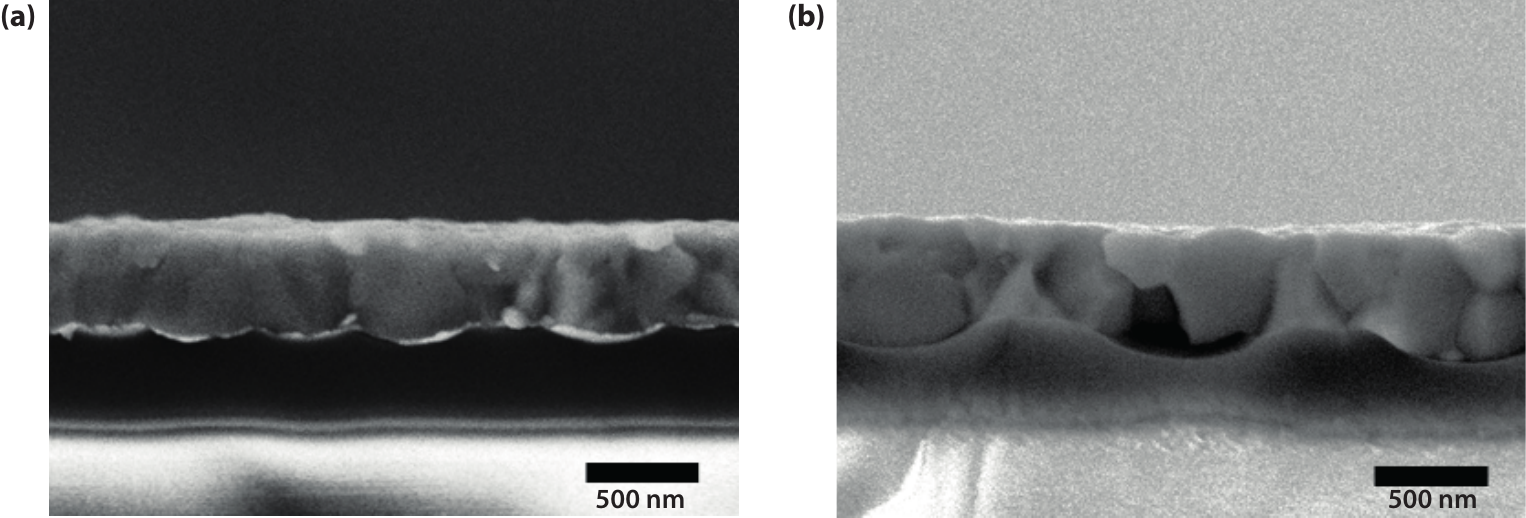}
\caption{5 kV SEM micrograph images of perovskite on sinusoidal substrates with (a) $P=500$~nm and $a=0.16$, (b) $P=750$~nm and $a=0.27$. The bright layer between substrate and perovskite represents the evaporated silver.}
\label{fig:SEMmicrographs}
\end{figure}

We investigate the morphology of the perovskite layers with scanning electron microscopy (SEM). Figures \ref{fig:SEMmicrographs}(a) and \ref{fig:SEMmicrographs}(b) show cross sectional SEM images of the perovskite on the two sinusoidal textures with $P=500(750)$~nm and $a=0.16(0.27)$. The perovskite films show thicknesses of 420-500~nm and 420-600~nm on the $P=500$~nm and $P=750$~nm nanotextured substrates, respectively. 
The layers are thicker than those required for good current matching, as we will discuss in Section \ref{sec:res} and Table \ref{table_d}. However, they can be easily adjusted by i.e. reducing the molarity of the perovskite precursor solution.

The perovskite fills the sinusoidal structures for both periodicities and can compensate a height difference of up to 200 nm on the substrate. On the top surface, the underlying sinusoidal nanotexture is not visible, which is in accordance with crystallization starting from the top surface. 

\begin{figure}
\centering
\includegraphics{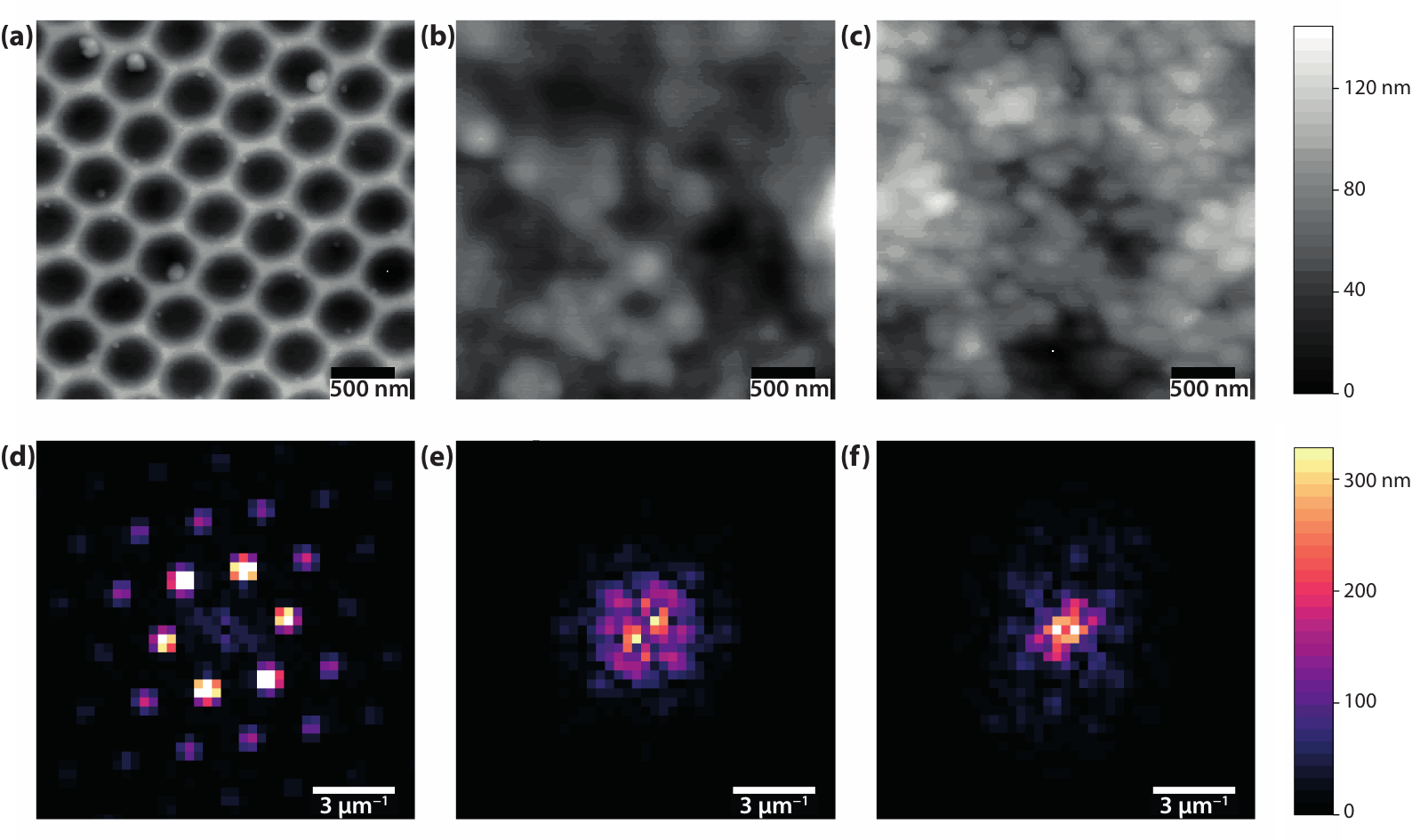}
 \caption{AFM micrograph images of (a) an uncoated, sinusoidally structured substrate, (b) perovskite on a structured substrate and (c) perovskite on flat PTAA as a reference. (d)-(f) display the 2D Fourier transforms corresponding to (a)-(c). }
\label{fig:AFMmicrographs}
\end{figure}

We further investigate the perovskite surface with atomic force microscopy (AFM) measurements.  Figure \ref{fig:AFMmicrographs}(a) displays a top view AFM micrograph (3~\textmu m$\times$ 3~\textmu m) of the bare nanotextured substrate with $P=500$~nm. The hexagonal arrangement of the sinusoidal structure is clearly visible. In contrast, the 3~\textmu m$\times$ 3~\textmu m micrographs in Fig. \ref{fig:AFMmicrographs} (b) and (c) show the disordered surfaces of perovskite layers on this structured substrate and on (flat) PTAA, respectively.  The perovskite films grown on these different substrates have a root mean square (RMS) roughness of 16~nm and 22~nm for the structured and flat substrates, respectively. The RMS roughness values of the perovskite layers in this study are larger than reported elsewhere \cite{jeon:2014,kim:2016}, which can be attributed to differences in compositions, processing and substrates.

Figures \ref{fig:AFMmicrographs}(d)-\ref{fig:AFMmicrographs}(f) show the \emph{Fourier transforms} (FT) corresponding to the AFM height profiles from Figs.~\ref{fig:AFMmicrographs}(a)-\ref{fig:AFMmicrographs}(c). The FT of the bare substrate features six hexagonally arranged spots of high intensity around the center and low intensity spots of higher order at larger wavenumbers. In contrast, the FTs of both perovskite layers in Figs. \ref{fig:AFMmicrographs}(e) and \ref{fig:AFMmicrographs}(f) feature a multitude of different wavevectors distributed over all angles. The highest amplitudes are at low frequencies at low wavenumbers, reflecting some unordered base signal which is modified by low amplitude, high frequency components to describe the small grains which were also found in Fig. \ref{fig:AFMmicrographs} (b) and (c). The FT of the perovskite layer on structured substrate [Fig.~\ref{fig:AFMmicrographs}(e)] resemblances that of the flat perovskite layer on PTAA [Fig.~\ref{fig:AFMmicrographs} (f)] much better that of the sinusoidal texture [Fig.~\ref{fig:AFMmicrographs}(a)]. Hence, the hexagonal texture of the substrate is not transfered to the perovskite top surface.

As we now have established that spin coating allows to fabricate perovskite films of a high morphological quality on sinusoidally nanotextured substrates, we can start with the optical simulations of nanotextured perovskite-silicon tandem solar cells.

\section{Simulation details}

\subsection{Simulated tandem cell architectures}
\begin{figure}
\centering
\includegraphics{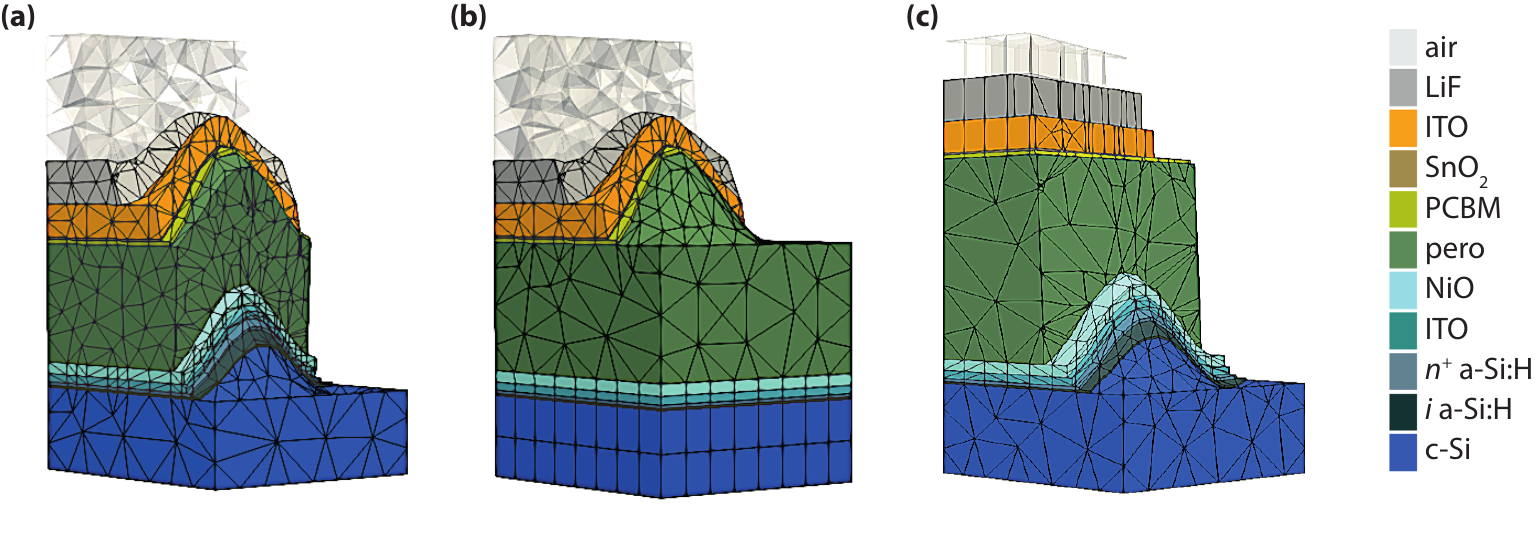}
\caption{Three different cell architectures with (a) double-side textured, (b) front-side textured and (c) back-side textured perovskite top cell. For the FEM simulations, the simulation domains are enclosed by air and silicon halfspaces, respectively. The light is incident from the air half space (top). The thicknesses of the layers are shown in Table \ref{table_v}.} 
\label{fig:layerstacks}
\end{figure}

For this study we numerically investigated the three architectures illustrated in Fig.~\ref{fig:layerstacks}: Figure \ref{fig:layerstacks}(a) shows the architecture with the \emph{double-side} textured perovskite top cell. In Figure \ref{fig:layerstacks}(b) the \emph{front side} of the perovskite absorber and all layers above are textured. In the \emph{rear-side} textured architecture, Fig.~\ref{fig:layerstacks}(c), only the intermediate layers between the Si absorber and the perovskite absorber are textured, while the interfaces on top of the perovskite layers remain flat. In all structures we assume conformal growth, meaning that all nanotextured interfaces have the same texture. For all architectures we use a period of $P = 500$~nm, which is very well suited for solar cells \cite{jaeger:2016opex,eisenhauer:2017, jaeger:2018opex}.

 As we have seen in Section \ref{sec:exp}, spin-coating allows to grow perovskite layers on nanotextured substrates. These layers are almost flat on top and hence resemble the architecture with the rear-side texture shown in Fig.~\ref{fig:layerstacks}(c). As demonstrated by other groups, a front-side textured architecture can be realised with \emph{nanoimprint} processes.\cite{paetzold:2015apl,brittman:2017,pourdavoud:2017} Paetzold and coworkers demonstrated that pattering the front surface does not affect the open circuit voltage $V_\text{oc}$ and fill factors of solar cells \cite{paetzold:2015apl}. One strategy to realize double-sided architectures is using \emph{vacuum deposition}, which can lead to conformal growth of the perovskite layer \cite{momblona:2016}.

In detail, the perovskite top cells illustrated in Fig.~\ref{fig:layerstacks} consist of a lithium fluoride (LiF) antireflective coating, an indium tin oxide (ITO) front contact, a tin oxide (\ce{SnO2}) buffer layer, an electron-selective phenyl-C61-butyric acid methyl ester (PCBM) contact layer, the perovskite absorber, a hole-selective nickel oxide (NiO) layer and the ITO interconnecting layer \cite{bush:2017, jaeger:2017pero}. The silicon bottom cell consists of $n^+$-doped  and intrinsic (i) hydrogenated amorphous silicon (a-Si:H) layers and the c-Si wafer absorber, which is assumed to be infinitely thick in the FEM simulations, as explained in Section  \ref{sec:sim-method}. All other layers take their thicknesses from previous work on optimizing a planar tandem device\cite{jaeger:2017eupvsec} and are displayed in Table \ref{table_v}.

\begin{table}
\centering
\sffamily
\begin{tabular}{lr} 
\hline
LiF & 107\\ 
ITO& 80 \\
\ce{SnO2} & 5\\ 
PCBM & 15\\
perovskite & 333\\
NiO & 10\\
ITO & 20\\
\textit{n}$^{+}$ a-Si:H & 8 \\
\textit{i} a-Si:H & 5 \\
 \hline
\end{tabular}
\rmfamily
\caption{The thicknesses of the layers in our simulated tandem cell. These are the set of optimal layer thicknesses numerically calculated in Ref.\ \citenum{jaeger:2017eupvsec}. The perovskite thickness has been adapted to take into account the different $(n,\,k)$ data used in this work. The air and c-Si layer are treated as infinite half spaces in the simulation. All layer thicknesses except the one of the perovskite are kept constant during the simulation. All values are in nm.}
\label{table_v}
\end{table}

\subsection{Mathematical description of hexagonal lattices}
The hexagonal sinusoidal nanotextures are mathematically described with 
\begin{equation} \label{eq:hex}
f_{\mathrm{hex}}(x,y) = -\frac{8}{9}\cos \left[\frac{1}{2}\left(x+\sqrt{3}y\right)\right]\cos\left[\frac{1}{2}\left(x-\sqrt{3}y\right)\right]\cos(x).
\end{equation}
We discussed them in detail in Ref.\ \citenum{jaeger:2016opex}, where the texture described in (\ref{eq:hex}) is referred to as \emph{negative cosine}. For the solar cells discussed in this work, we also tested the \emph{positive cosine} textures and just as in Ref.\ \citenum{jaeger:2016opex} they are outperformed by the negative cosine texture. Therefore, we only report about solar cells with negative cosine textures in this manuscript. 

We can scale the nanotexture to the desired period $P$, which is the side length of the rhombus-shaped unit cell, with the following substitutions
\begin{equation} \label{eq:sub}
x \rightarrow \frac{2\pi}{\sqrt{3}P}x
\quad\text{and}\quad 
y \rightarrow \frac{2\pi}{\sqrt{3}P}y .
\end{equation}
The valley-to-peak height $h$ of the nanotexture can be set by multiplying (\ref{eq:hex}) with $h$. The aspect ratio $a$ is defined as $a=h/P$; small or large values of $a$ lead to flat or steep textures, respectively. 

\subsection{Simulation methods}
\label{sec:sim-method}
The optical simulations are performed with the finite-element method (FEM) solver \texttt{JCMsuite} \cite{pomplun:2007}, which provides a rigorous solution to Maxwell's equations for a given structure. As illustrated in Fig.~\ref{fig:layerstacks} the 3-dimensional structures are meshed using tetrahedral and prismoidal elements.

The complex refractive index spectra ($n$, $k$) used for the simulations were determined as follows: perovskite data were retrieved using ellipsometry and transmittance/reflectance spectrophotometry \cite{guerra:2017}. For \ce{NiO_x} \cite{you:2016} and the sputtered ITO layers, ellipsometry and the program RIGVM was used \cite{pflug:2004}. The data for PCBM was extracted from reflectance/transmittance measurements with the method described in Refs.\ \citenum{albrecht:2014phd, djurisic:2000}. The \ce{SnO2} layers were deposited using plasma-enhanced atomic layer deposition and characterized with ellipsometry \cite{chistiakova:2018}. The data for the RF-PECVD a-Si:H layers\cite{mazzarella:2017} were extracted using SCOUT \cite{scout}. For LiF \cite{li:1976} and spiro-OMeTAD\cite{filipic:2015} we used data from literature. 

We set the side-length constraint of the three-dimensional elements (prisms and tetrahedrals) for the FEM simulations to $\lambda/n(\lambda)$, where $\lambda$ is the incident wavelength and $n(\lambda)$ is the refractive index of the material. In this way one grid was generated for every wavelength interval of 100~nm width.  Employing higher polynomial degrees (between 3 and 6) rather than finer meshes in an adaptive  $hp$-FEM implementation allows to achieve accurate results at optimal computational costs\cite{burger:2015}. A convergence study confirmed the accuracy of the  pre-set accuracy value.  To accurately reconstruct the sinusoidal interfaces, we set their surface side-length constraint to between 55 and 17~nm, depending on the aspect ratio. To obtain the angular dependence of the hexagonal unit cell, the incident angle is varied along the 0$^{\circ}$ and 30$^{\circ}$ axes and an average is taken. \cite{jaeger:2016opex}

We model the c-Si absorber and the air layer as infinite half spaces, which is numerically realized by using perfectly matched layers (PML) on the top and bottom boundaries. For the other faces of the computational domain periodic boundary conditions are applied. Light is incident from the air half space.

\begin{figure}
\centering
 \includegraphics{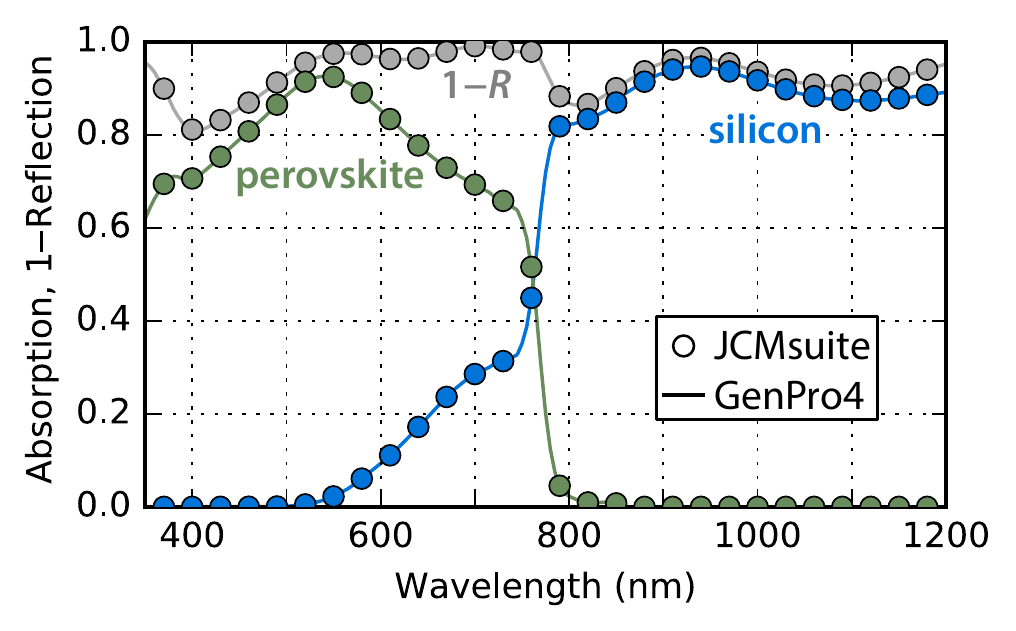}
 \setlength{\abovecaptionskip}{6pt}
 \caption{Absorption spectra of the c-Si (blue) and the perovskite (green) layers and one minus the overall reflection (grey) for a flat reference structure with a flat layer stack with the layers as in Table~\ref{table_v} with an infinitely-thick silicon layer. Results were computed with the FEM solver \texttt{JCMsuite} (circles) and the analytical tool \texttt{GenPro4} (lines).}
 \label{fig:sim_spec}	
\end{figure}

To cross-check the accuracy of the simulations we compared simulations for the planar tandem device with \texttt{JCMsuite} (3D FEM) and \texttt{GenPro4}, which can combine wave optics for coherent thin layers and ray optics for thick incoherent layers.\cite{santbergen:2017} As can be seen in Fig.~\ref{fig:sim_spec}, the results obtained with the two methods are in excellent agreement. 

However, the Si absorber of our tandem cell has a realistic thickness of 160~\textmu m. In order to account for this we approximate the absorption in the Si absorber by assuming that the rear side of the Si behaves as a \emph{Lambertian} scatterer. In practice this means that the wavelength dependent absorbtion coefficient of the infinite Si layer $\alpha(\lambda)$---which is obtained using the light transmitted into the Si in the FEM simulation---is converted to the absorption of a finite layer $A(\lambda)$ with \emph{Lambertian} scattering using the the Tiedje-Yablonovitch limit

\begin{equation}
\label{eq:a}
A(\lambda) = \frac{\alpha(\lambda)}{\alpha(\lambda)+\left(4 \left[n(\lambda)\right]^2L\right)^{-1}},	
\end{equation}
which is a function of the absorber thickness $L$, the wavelength dependent absorption coefficient $\alpha(\lambda)$ and the wavelength dependent refractive index $n(\lambda)$ \cite{tiedje:1984}.

As a figure of merit we use the photocurrent density $J_{\mathrm{ph}}$, which can be calculated for the $i$-th layer from the absorption profile using 
\begin{equation} \label{eq:j_ph}
J_{\mathrm{ph},i} = -e\int_{310~\mathrm{nm}}^{1200~\mathrm{nm}} A_{i}(\lambda)\Phi_{AM1.5}(\lambda)\cos\theta_{in}\mathrm{d}\lambda, 	
\end{equation}
where $e$ is the elementary charge, $A_{i}(\lambda)$ the absorption spectrum of the $i$-th layer, $\Phi_{AM1.5}(\lambda)$ the spectral photon flux under AM1.5G condition \cite{iec:60904-3} and $\theta_{in}$ the angle between the incident light and the solar cell normal. The photocurrent densities calculated for the perovskite and silicon layers are the \emph{maximum achievable current densities}, because we assume that all the absorbed light leads to the generation of electron-hole pairs, which can be extracted from the solar cell. The current densities for the other layers are losses due to parasitic absorption.

In order to estimate the \emph{potential power conversion efficiency} of the tandem device the open circuit voltage ($V_\text{oc}$) of each subcell and the fill factor (FF) of the overall device are required. Additionally the logarithmic dependence of the $V_\text{oc}$ on the photocurrent density must be taken into account. We assume the fill factor to be 81\% \cite{albrecht:2016}. Furthermore we assume that a single junction perovskite and Si solar cell under standard illumination conditions have a short circuit current density of 22~mA/cm$^{2}$ and 42~mA/cm$^{2}$ with a related open circuit voltage of 1.130~V and 0.730~V, respectively. These values will then be used to estimate the $V_\text{oc}$ values for the sub-cells in the tandem device.

In order to determine the perovskite thickness for current matching, we use Newton's method. The method begins at a starting value $x_0$ and  repeats the iteration
\begin{equation} \label{eq:newton}
x_{n+1} = x_n - \frac{f(x_n)}{f'(x_n)}	
\end{equation}
until a sufficiently accurate value is reached. 
The partial derivatives of the absorption of the perovskite and c-Si absorber layers with respect to the perovskite layer thickness are computed directly in \texttt{JCMsuite} \cite{burger:2013derivative}.

\begin{figure}
\centering
  \includegraphics{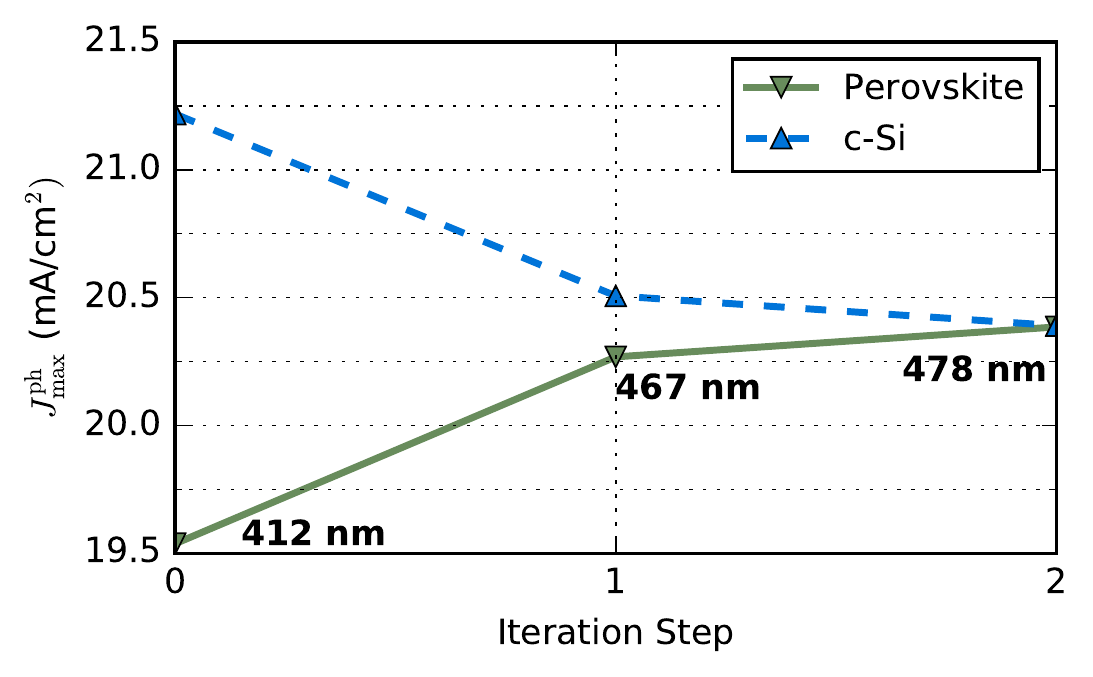}
  \caption{Example for applying Newton's method to adapt the perovskite thickness for current matching. For the back side textured structure with an aspect ratio of $a=1.0$. The effective perovskite thickness after each iteration step using Newton's method is given; only two iterations are required to reach current matching.} 
\label{fig:newton}
\end{figure}

As starting value for the architecture with the smallest aspect ratio we choose a perovskite thickness of 333~nm, which is the optimal value for the planar architecture \cite{jaeger:2017eupvsec}. For each new value of aspect ratio we use as starting value the optimized value for the previous aspect ratio. By employing this method we reduce the number of simulations, because no extensive thickness parameter scans are required. An example is shown in Fig.~\ref{fig:newton}, where we see the current densities in the perovskite and silicon subcells after each iteration for the back side textured structure with aspect ratio $a=1.0$. To reach current matching, the perovskite thickness had to be increased from 412~nm to 478~nm.

\section{Simulation results}
\label{sec:res}

\begin{figure}
\centering
 \includegraphics{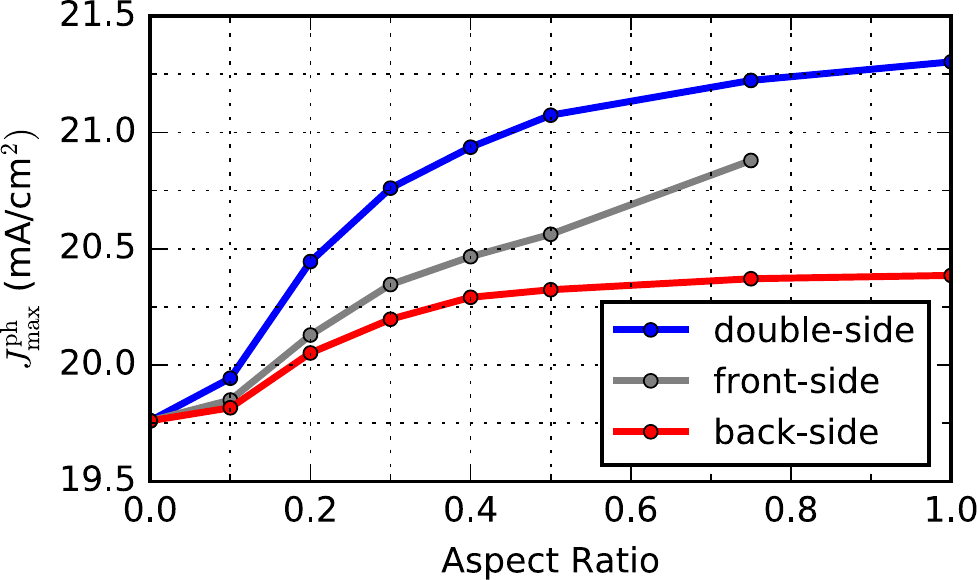}
 \caption{The maximum achievable matched current density $J_\text{ph}^\text{max}$ after thickness optimization shown in dependence of the aspect ratio for the three architectures shown in \ref{fig:layerstacks}. The corresponding optimal effective perovskite layer thicknesses are shown in Table \ref{table_d}.}
 \label{fig:all_struc}	
\end{figure}

Figure  \ref{fig:all_struc} shows the matched maximum achievable current density as a function of the aspect ratio $a$ for the three architectures illustrated in Fig. \ref{fig:layerstacks}. For all three textures, the largest aspect ratio leads to the the maximum current densities, which are 21.3~mA/cm$^2$, 20.9~mA/cm$^2$ and 20.4~mA/cm$^2$ for the double, front and back sided textures, respectively. For the double- and back-side textures that was $a$ = 1.0, while for the front-side texture that was $a$ = 0.75. An aspect ratio of 1.0 was not achievable with this texture since the layer thickness required for current matching would be smaller than the texture height. As discussed in section \ref{sec:exp}, the  architecture with the back-side textured solar cell (Fig.~\ref{fig:all_struc}, red lines), is closest to the experimentally realizable architecture. Note that in this case the difference in the current densities $J^\text{max}_\text{ph}$ between $a$ = 0.3, 0.4 and 0.5 is less than 0.1~mA/cm$^2$. 

The effective perovskite thicknesses $d$ at current match conditions are shown in Table \ref{table_d}; $d$ is the thickness a flat perovskite layer would have with the same volume and hence the same amount of material. The values are obtained by dividing the volumes of the perovskite absorber by the surface of the base. Comparing the effective thicknesses with the associated current densities we can see that a high effective thickness does not necessarily relate to a high absorption rate. For the architecture where only the layers above the perovskite absorber are textured (front-side) we find the highest proportion of absorption rate in the perovskite layer per effective thickness for all aspect ratios.

\begin{table}
\centering
\sffamily
\begin{tabular}{llllllll}
 \hline

  \textbf{aspect ratio} & \textbf{0.1} & \textbf{0.2} & \textbf{0.3} & \textbf{0.4} & \textbf{0.5} & \textbf{0.75} & \textbf{1.0}\\
 \hline
double-side & 333 & 340 & 343 & 350 & 360 & 378 & 386\\ 
front-side  & 324 & 306 & 283 & 262 & 243 & 201& -\\
back-side   & 338 & 364 & 386 & 402 & 411 & 439& 478\\
\hline
\end{tabular}
\setlength{\abovecaptionskip}{10pt}
\rmfamily
\caption{The effective perovskite thicknesses of the three different architectures illustrated in Fig.\ \ref{fig:layerstacks} with various aspect ratios after thickness optimization for current matching. All values are in nm.}
\label{table_d}
\end{table}

Starting from the photocurrent density and $V_\text{oc}$ of single junction cells given in Section \ref{sec:sim-method}, the matched photocurrent density values for the planar tandem of 19.7~mA/cm$^{2}$ and 19.8~mA/cm$^{2}$ results in $V_\text{oc}$ values of 1.127~V and 0.711~V for the perovskite and Si sub-cells, respectively. This leads to a power conversion efficiency of 29.3\%. The matched photo-current density for device with a double sided texture with $a$= 1.0 is 21.3~mA/cm$^{2}$, which results in $V_\text{oc}$ values of 1.129~V and 0.713~V for the perovskite and Si sub-cells, respectively, leading to a power conversion efficiency of 31.8\%. This represents an increase of 2.5\% efficiency absolute at the standard perovskite bandgap of 1.56~eV. For the realistic back-textured architecture the matched current density is 20.3~mA/cm$^{2}$ at $a$ = 0.5, leading to V$_{oc}$ values of 1.128~V and 0.711~V for the perovskite and Si cells respectively. That results in an overall power conversion efficiency of 30.2\%.

\begin{figure}
\centering
\includegraphics{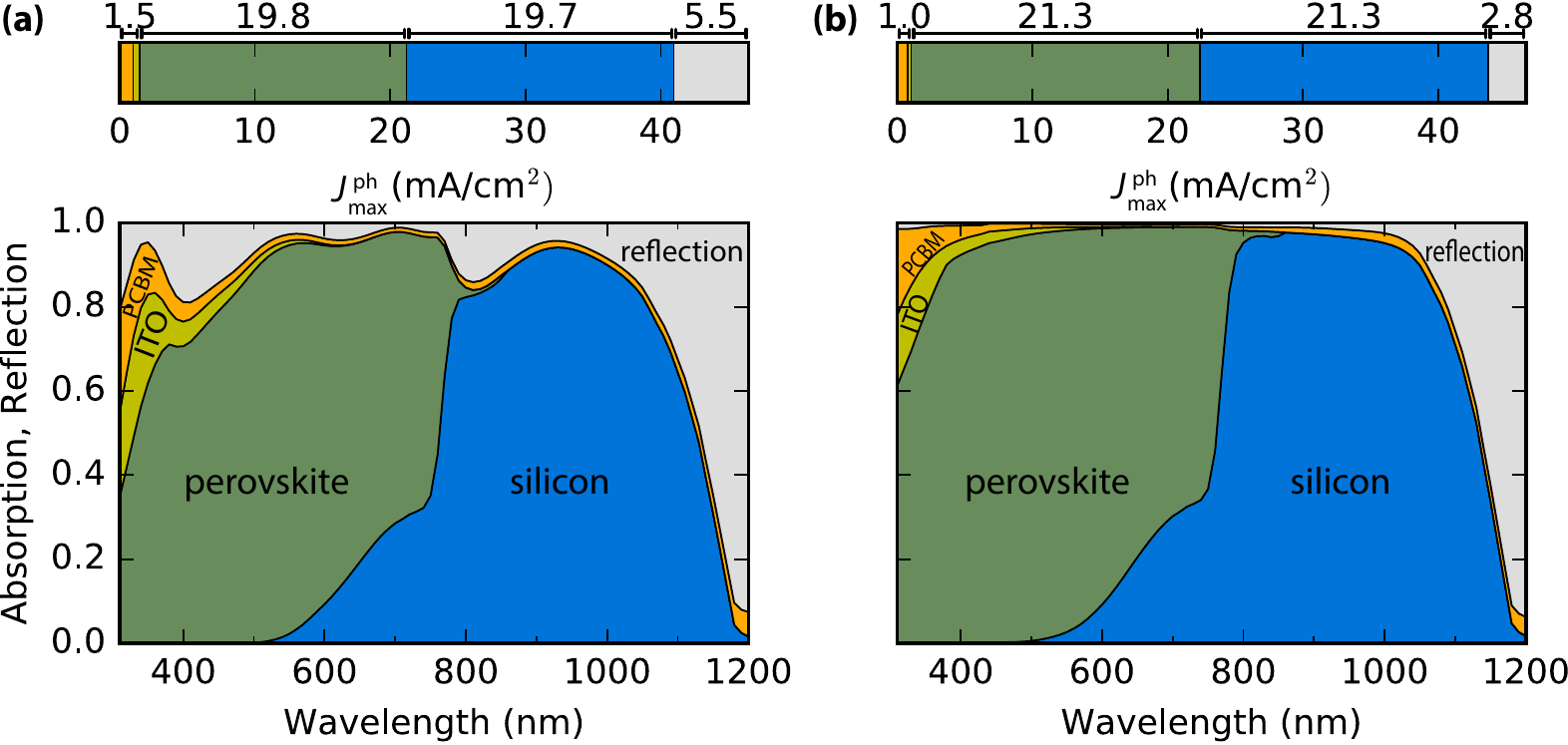}
 \caption{Absorption profile and associated theoretical current density for the (a) planar and (b) fully textured architecture with $a$ = 1.0. With a maximum achievable current density of 21.3\,mA/cm$^2$ for each absorber cell we have a power conversion efficiency of 31.8\%.}
 \label{fig:abs_prof}	
\end{figure}

Figure \ref{fig:abs_prof} shows the absorption profile for silicon perovskite solar cells with (a) a planar and (b) a fully textured perovskite top cell with aspect ratio $a$ =1.0. Additionally the corresponding current densities are shown. Note that only the current densities from the perovskite and silicon layers can contribute to electricity generation, while the other values are parasitic absorption or reflection losses. For silicon, we assume \emph{Lambertian} scattering at the back by using Eq.\ \ref{eq:a}. Compared to the planar architecture an additional current density of about 1.5~mA/cm$^2$ per subcell can be collected by implementing the double sided texture with $a$ = 1.0. 

We see that the planar device suffers from large reflective losses in the wavelength range from 400 to 1100~nm. These losses oscillate in magnitude which is typical for Fabry-Perot type reflections from a planar thin film layer stack. In contrast, the fully textured device shows negligible reflection losses in the same wavelength region. Hence, the overall reflection losses are reduced by 2.7~mA/cm$^2$. The sinusoidal texture scatters the light, thereby stopping multiple reflections necessary for Fabry-Perot oscillations, which is evident from the reflection curve being largely independent of wavelength. At the same time the texture reduces the total reflection which directly increases the absorption in the absorber layers.

\begin{figure}
\centering
\includegraphics{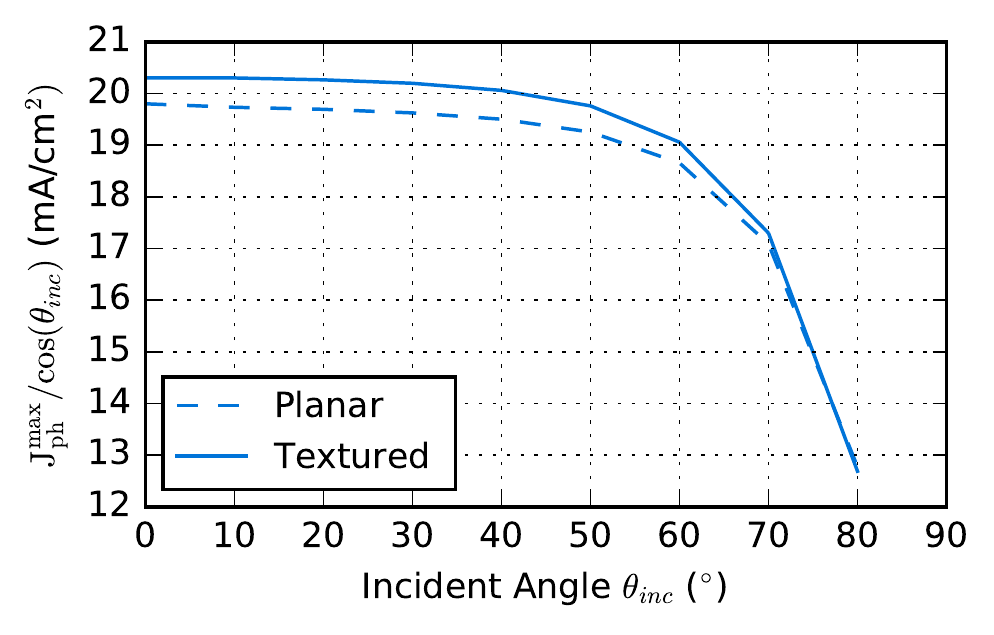}
 \caption{The angular dependence of the maximum matched current density for the planar and back textured ($a$=0.5) structure.}
 \label{fig:ang_dep}	
\end{figure}

The angular dependent performance is also an important factor for solar cells without active tracking. Figure \ref{fig:ang_dep} shows how the maximum matched current density varies with the angle of incidence for both the optimized planar device and the back textured device with aspect ratio 0.5. Since the incident power falls off as $\cos(\theta_{inc})$ where $\theta_{inc}$ is the incident angle, we divide the maximum matched current density by this factor in order to highlight the angular dependent difference in light absorption between the planar and back textured device, independent of the incident power. It can be seen that the advantage in absorption and therefore current density provided by the nanotexture is maintained over a wide angle range up to 60$^{\circ}$. This supports the application of sinusoidal nanotextures for solar applications.

\section{Conclusion}
In this work we numerically studied \emph{nanophotonic light management} concepts for perovskite-silicon tandem solar cells. As nanophotonic structures we used hexagonal sinusoidal nanotextures. For current matching, we optimized the thickness of the perovskite absorber with Newton's method, which quickly converged after three iteration steps at most. To experimentally motivate this numerical study, we demonstrated the deposition of perovskite layers onto sinusoidally structured nanophotonic substrates with spin coating.

The highest maximum achievable photocurrent density $J^\text{max}_\text{ph}$ = 21.3~mA/cm$^2$ in each subcell was obtained with the fully textured structure with aspect ratio $a$ = 1.0, which corresponds to a power conversion efficiency of 31.8\% at the standard perovskite bandgap. Compared to the planar architecture we have an increase in $J^\text{max}_\text{ph}$ of 1.6~mA/cm$^2$ for each absorber layer and a potential increase of power conversion efficiency of 2.5\% absolute. The efficiency can be increased even further by using perovskite materials with higher bandgaps, where the optimal bandgap is around 1.7~eV \cite{albrecht:2016,jaeger:2017pero}.

Future experimental efforts should focus on the implementation of double-side textured perovskite layers onto high-aspect-ratio substrates as well as to the experimental realization of textured perovskite single-junctions. Special care has to be taken to investigate the implications of structured substrates for perovskite crystal quality and thus electrical performance parameters which were neglected in this study. Ultimately, future efforts should be directed to fully textured and highly efficient perovskite-silicon tandem solar cells.

\subsection*{Experimental Details}
The PTAA layer for the perovskite reference sample is prepared by spin-coating 100\,\textmu L of a 2\,mg PTAA (Aldrich) per mL of Toluene (Aldrich) solution onto an ITO-coated substrate (Automatic Research). After spin-coating at 4000\,rpm for 30\,s, the sample is annealed at 100\,\textdegree C for 10\,min.

For the mixed cation, mixed halide perovskite, two 1.5~M precursor solutions of \ce{PbI2} (TCI) and \ce{PbBr2} (TCI) in dimethylformamide (DMF, Aldrich) and dimethyl sulfoxide (DMSO, Aldrich) (4:1 v:v) are mixed with formamidinium iodide (FAI, Dyenamo) and methylammonium bromide (MABr, Dyenamo), respectively. The resulting \ce{FAPbI3} and \ce{MAPbBr3} solutions with a Pb-to-cation ratio of 1.09:1 are mixed in a 5:1 ($v$:$v$) ratio and supplemented by addition of 5 vol\% of a  1.5~cesium iodide (CsI, abcr) in DMSO solution. This solution is then diluted down to a 0.8~M solution to obtain films in the desired thickness range. For this, DMF:DMSO 4:1 ($v$:$v$) is added to the as-prepared solution.

For fabricating the perovskite layers, the antisolvent method was used. For this, the perovskite precursor solution was spin-coated onto the different substrates at 1000 and 6000\,rpm for 10 and 20\,s, respectively. 5\,s prior to the end of process, 200\,\textmu L of chlorobenzene (Aldrich) was dripped as antisolvent. After this, the sample was directly transferred to a hot plate and baked at 100\,\textdegree C for 1\,hour.

\subsection*{Disclosures}
The authors have no relevant financial interests in the manuscript and no other potential conflicts of interest to disclose.

\acknowledgments 

The numerical results were obtained at the \emph{Berlin Joint Lab for Optical Simulations for Energy Research} (BerOSE) of Helmholtz-Zentrum Berlin f\"{u}r Materialien und Energie, Zuse Institute Berlin and Freie Universit\"{a}t Berlin. 

D.E., G. K., C.B. and K.J. acknowledge the German Federal Ministry of Education and Research (BMBF) for funding the research activities of the Nano-SIPPE group within the program NanoMatFutur (grant no. 03X5520). 
P.M. is funded by the Helmholtz Innovation Lab HySPRINT, which is financially supported by the Helmholtz Association.
M.H. and S.B. acknowledge support by Einstein Foundation Berlin through ECMath within subproject~SE6.
S.A. acknowledges the BMBF within the project “Materialforschung f\"{u}r die Energiewende” for funding of his Young Investigator Group (grant no. 03SF0540);  together with P.T. he acknowledges the German Federal Ministry for Economic Affairs and Energy (BMWi) for funding of the “PersiST” project (grant no. 0324037C).

We thank Johannes Sutter for conducting the AFM measurements and Florian Ruske for supporting us with the determination of optical properties.


\end{document}